\RequirePackage{fix-cm}
\documentclass[twocolumn]{svjour3}          % twocolumn
\smartqed  % flush right qed marks, e.g. at end of proof
\usepackage{graphicx}
\usepackage{hyperref}
\usepackage{multirow}
\usepackage{makecell}
\usepackage[dvipsnames]{xcolor}
\begin{document}

\title{CUBES Phase A design overview
}
\subtitle{The Cassegrain U-Band Efficient Spectrograph for the Very Large Telescope} 
%at ESO's Paranal Observatory.}


\author{Alessio Zanutta\and
Stefano Cristiani\and
David Atkinson\and
Veronica Baldini\and
Andrea Balestra \and
Beatriz Barbuy\and
Vanessa Bawden P. Macanhan\and
Ariadna Calcines\and
Giorgio Calderone \and
Scott Case\and
Bruno V. Castilho\and
Gabriele Cescutti\and
Roberto Cirami\and
Igor Coretti\and
Stefano Covino\and
Guido Cupani\and
Vincenzo De Caprio\and
Hans Dekker\and
Paolo Di Marcantonio\and
Valentina D'Odorico\and
%Simon Ellis\and
Heitor Ernandes\and
Chris Evans\and
Tobias Feger\and
Carmen Feiz\and
Mariagrazia Franchini\and
Matteo Genoni\and
Clemens D. Gneiding\and
Mikołaj Kałuszyński\and
Marco Landoni\and
Jon Lawrence\and
David Lunney\and
Chris Miller\and
%Mahesh Mohanan\and
Karan Molaverdikhani\and
%Kieran O'Brien\and
Cyrielle Opitom\and
Giorgio Pariani\and
Silvia Piranomonte\and
Andreas Quirrenbach\and
Edoardo Maria Alberto Redaelli\and
Marco Riva\and
David Robertson\and
Silvia Rossi\and
Florian Rothmaier\and
Walter Seifert\and
Rodolfo Smiljanic\and
Julian St\"urmer\and
Ingo Stilz\and
Andrea Trost\and
Orlando Verducci\and
Chris Waring\and
Stephen Watson\and
Martyn Wells\and
Wenli Xu\and
Tayyaba Zafar\and
Sonia Zorba \and
%Jessica Zheng
%etc.
}


\institute{
            A.~Zanutta, M.~Riva, M.~Genoni, G.~Pariani, S.~Covino, M.~Landoni \& E.M.A.~Redaelli \at
            INAF - Osservatorio Astronomico di Brera,
            via E. Bianchi 46, 23807, Merate (LC), Italy\\
            \email{alessio.zanutta@inaf.it}\\
            % -----------------------------------------
            \and
            S.~Cristiani, V.~Baldini, G.~Calderone, G.~Cescutti, R.~Cirami, I.~Coretti, G.~Cupani, P.~Di Marcantonio, V.~D'Odorico, M.~Franchini \& S.~Zorba \at
            INAF - Osservatorio Astronomico di Trieste, via G.B. Tiepolo 11, 34143, Trieste, Italy\\
            % -----------------------------------------
            \and
            D.~Atkinson, C.~Evans, D.~Lunney, C.~Miller, C.~Waring, S.~Watson \& M.~Wells \at
            UK Astronomy Technology Centre, Royal Observatory, Blackford Hill, Edinburgh EH9 3HJ, UK\\
            % -----------------------------------------
            \and
            B.~Barbuy, H.~Ernandes \& S.~Rossi \at
            IAG, Universidade de S\~ao Paulo, Rua do Mat\~ao 1226, S\~ao Paulo 05508-090, Brazil\\
            % -----------------------------------------
            \and
            B.V.~Castilho, V.B.P.~Macanham, C.D.~Gneiding \& O.~Verducci \at
            Laboratório Nacional de Astrofísica, MCTI, Rua Estados Unidos 154, Itajubá, 37504364, MG, Brazil\\
            % -----------------------------------------
            \and
            V. De Caprio \at
            INAF - Osservatorio Astronomico di Capodimonte, Salita Moiariello 16, 80131, Napoli, Italy\\
            % -----------------------------------------
            \and
            A.~Balestra \at
            INAF - Osservatorio Astronomico di Padova, Vicolo dell'Osservatorio, 5, 35122, Padova, Italy\\
            % -----------------------------------------
            \and
            A.~Calcines \at
            Durham University, Centre for Advanced Instrumentation, Durham, UK\\
            % -----------------------------------------
            \and
            S.~Case, J.~Lawrence, D.~Robertson, T.~Zafar, \& T.~Feger \at
            Australian Astronomical Optics - Macquarie, Macquarie University, NSW 2109, Australia\\
            % -----------------------------------------
            \and
            M.~Landoni \at
            INAF - Osservatorio Astronomico di Cagliari, via della Scienza 5, 09047 Selargius (CA), Italy\\
            % -----------------------------------------
            \and
            S.~Piranomonte \at
            INAF - Osservatorio Astronomico di Roma, Via Frascati 33, I-00040 Monte Porzio Catone (RM), 00078, Italy\\
            % -----------------------------------------
            \and
            R.~Smiljanic \& M.~Kałuszyński \at
            Nicolaus Copernicus Astronomical Center, Polish Academy of Sciences, ul. Bartycka 18, 00-716, Warsaw, Poland\\
            % -----------------------------------------
            \and
            W.~Seifert, F.~Rothmaier, J.~St\"urmer, I.~Stilz, C.~Feiz, K.~Molaverdikhani \& A.~Quirrenbach \at
            Landessternwarte, Zentrum für Astronomie der Universität Heidelberg, Königstuhl 12, 69117, Heidelberg, Germany\\
            % -----------------------------------------
            \and
            H.~Dekker \at
            Consultant Astronomical Instrumentation, Alpenrosenstr. 15, 85521 Ottobrunn, Germany\\
            % -----------------------------------------
            \and
            W.~Xu \at
            Optical System Engineering (OSE), Kirchenstr. 6, 74937, Spechbach, Germany\\
            % -----------------------------------------
            \and
            C.~Opitom \at
            Institute for Astronomy, University of Edinburgh, Royal Observatory, Edinburgh, EH9 3HJ, UK\\
            % -----------------------------------------
            \and
            K.~Molaverdikhani \at
            Max-Planck-Institut für Astronomie, Königstuhl 17, D-69117 Heidelberg, Germany 
            \& Universitäts-Sternwarte, Ludwig-Maximilians-Universität München, Scheinerstrasse 1, D-81679 München, Germany \\
            % -----------------------------------------
            \and
            A.~Trost \at
            Department of Physics, University of Trieste, Via A. Valerio 2, 34127 Trieste, Italy\\
}

\date{Received: date / Accepted: date}

\maketitle

\begin{abstract}
We present the baseline conceptual design of the Cassegrain U-Band Efficient Spectrograph (CUBES) for the Very Large Telescope. CUBES will provide unprecedented sensitivity
for spectroscopy on a 8\,--\,10\,m class telescope in the ground ultraviolet (UV), spanning a
bandwidth of $\ge$100\,nm that starts at 300\,nm, the shortest wavelength accessible from the ground.
%In this paper we present the baseline design overview of the spectrograph CUBES (Cassegrain U-Band Efficient Spectrograph) developed during the Phase-A study for the Very Large Telescope (VLT).
The design has been optimized for end-to-end efficiency and provides a spectral resolving power of $R$\,$\ge$\,20000, that will unlock a broad range of new topics across solar system, Galactic and extraglactic astronomy. The design also features a second, lower-resolution ($R$\,$\sim$\,7000) mode and has the option of a fiberlink to the UVES instrument for simultaneous observations at longer wavelengths.

%In this paper, after an introduction of the scientific context, 
Here we present the optical, mechanical and software design of the various subsystems of the instrument after the Phase A study of the project. We discuss the expected performances for the layout choices and highlight some of the performance trade-offs considered to best meet the instrument top-level requirements. We also introduce the model-based system engineering approach used to organize and manage the project activities and interfaces, in the context that it is increasingly necessary to integrate such tools in the development of complex
%effectively handle 
astronomical projects.

\end{abstract}

\section{Introduction}
\label{intro}
High-resolution ($R$\,$>$\,20000) spectroscopy in the near-UV regime provides access to a very powerful and relevant observational window that has not yet been deeply explored from the ground. The current state-of-the art facility at the European Southern Observatory (ESO) for high-resolution spectroscopy at wavelengths shortwards of 400\,nm is the Ultraviolet and Visual Echelle Spectrograph (UVES) \cite{Ref-UVES} on the Very Large Telescope (VLT), but its efficiency is limited to just a few percent in the near-UV. This means that across a broad range of different scientific fields, observations are limited to small, bright samples of objects. 

The ground UV is an immensely rich and important part of the spectrum for exciting contemporary topics in solar system, Galactic and extragalactic astronomy and astrophysics (for background see e.g. \cite{2018SPIE10702E..2EE,2019vltt.confE..52S,2020SPIE11447E..60E,Barb_2014}). These range from searching for water in the asteroid belt, to addressing significant questions regarding stellar nucleosynthesis of iron-peak and heavy elements, and some lighter elements (notably Beryllium), to fundamental studies of the primordial deuterium abundance and the contribution of galaxies to the cosmic UV background.
%For instance, it allows to detect signature ascribed to a tremendous diversity of iron-peak and heavy elements in stellar spectra, as well as to some lighter elements (notably Beryllium) and light-element molecules (CO, CN, OH) opening the possibility to study stellar evolution and element formation with unprecedented details. From the extragalatic point of view, observations in the UV regime allow to deeply investigate absorption signatures in the circumgalactic medium (CGM) of distant galaxies, a proxy for measuring the contribution of different types of galaxies or Active Galactic Nuclei (AGN) to the cosmic UV background 

Given the strong scientific motivations to improve spectroscopic sensitivity at the shortest wavelengths accessible from the ground, ESO issued a call for a new UV spectrograph for the VLT, to provide a significant gain in throughput compared to current facilities. Our assembled consortium responded to that call with the concept of the Cassegrain U-Band Efficient Spectrograph (CUBES).

In the resulting conceptual design (Phase~A) study we developed a design for a Cassegrain spectrograph that provides excellent throughput and a spectral resolving power of $R\,\ge\,20000$ over 300\,--\,405\,nm. Many of the assembled science cases would also benefit from simultaneous spectroscopy at longer wavelengths, so the Phase~A design also includes an (optional) fiber-feed system to UVES.
%To exploit the capabilities offered by UVES for some science targets, we also designed an optional fiber-feed system for simultaneous observations at longer wavelengths. 

Combining its unprecedented performance and spectral resolution, CUBES will become the UV workhorse for the VLT and, given its competitiveness at short wavelengths, will be the blue-eye of the sky alongside the Extremely Large Telescope (ELT), providing a world-leading capability for ESO well into the 2030s. 

Here we present the results and current design of CUBES from the Phase~A study. %In particular, we focus on the optical and mechanical system layout and we discuss its performance against the design choices made during the study. 
This paper is organized as follows. In Sects~2 and 3 we detail the optical, mechanical and software design of CUBES and we give a brief overview of the consortium and construction schedule for the instrument in Sect.~4. We present the predicted instrument performance in Sect.~5 and introduce our model-based system engineering approach, adopted from the very beginning, in Sect.~6. Concluding remarks are given in Sect.~7.

\section{The CUBES instrument: requirements and sub-systems description}
\subsection{Scientific motivations}

A broad range of motivating cases for high-sensitivity spectroscopy in the ground UV were developed by the CUBES Science Team during the Phase~A study. An overview of the science case and the specific cases which drive the technical requirements of the instrument is given elsewhere in this Special Issue \cite{ExA_Sci_Overview}. To illustrate the breadth of the assembled cases, we also refer the reader to the more detailed articles on cometary science \cite{ExA_Opitom}, accretion and outflows in young stellar objects (YSOs) \cite{ExA_Alcala}, metal-poor stars \cite{ExA_Bonifacio,ExA_Hansen,ExA_Ernandes}, beryllium abundances in globular clusters \cite{ExA_Giribaldi}, extragalactic massive stars \cite{ExA_Evans}, tracing the baryonic mass in the high-redshift circumgalactic medium \cite{ExA_DOdorico}, and studies of molecular hydrogen at high redshift \cite{ExA_Noterdaeme}.

The cases developed for studies of transients with CUBES also require simultaneous spectroscopic coverage at longer (visible) wavelengths, e.g. with UVES
(see \cite{ExA_Sci_Overview}). Similarly, some of the observations envisaged for studies of YSOs also require simultaneous spectroscopy with UVES \cite{ExA_Alcala}. In the Phase~A design we therefore also explored the option of a fiber-link to UVES to enable simultaneous spectroscopy of the CUBES targets at $\lambda$\,$>$\,410\,nm. Simultaneous longer-wavelength observations are required for the transients and YSOs for astrophysical reasons, but we note that approximately half of the other cases developed for CUBES will benefit from simultaneous UVES observations, as they will provide complementary/supporting data more efficiently than observing separately with the two instruments.

\subsection{Top Level Requirements (TLRs)}
\label{sec:req} 
%A broad range of scientific cases that require better spectroscopic sensitivity in the near UV were developed during the Phase A study, some of which are outlined in detail in other papers in this Special Issue. 
We used the assembled science cases to drive the design concept by identifying the TLRs that the instrument shall provide, keeping in mind that these impose limits on the design choices in terms of the delivered resolving power ($R$), efficiency, wavelength range, etc. Key TLRs identified were:
\begin{itemize}
\item \textbf{Spectral range:} CUBES shall provide a spectrum of the target over the entire wavelength range of 305\,--\,400\,nm in a single exposure (goal: 300\,--\,420\,nm).
\item\textbf{Efficiency:} The efficiency of the spectrograph, from slit to detector (included), shall be $>$40\% for 305\,--\,360\,nm (goal $>$45\%, with $>$50\% at 313\,nm), and $>$37\% (goal 40\%) between 360 and 400\,nm.
\item \textbf{Resolving power (\textit{R}):} In any part of the spectrum, $R$ shall be $>$19000, with an average value $>$20000. $R$ is defined $R=\lambda/d\lambda$ where $d\lambda$ is the full width at half maximum (FWHM) of unresolved spectral lines of a hollow cathode lamp uniformly illuminating the CUBES entrance aperture.
\item \textbf{Signal-to-noise (S/N) ratio:} In a 1\,hr exposure the spectrograph shall be able to obtain, for an A0-type star of $U$\,$=$\,17.5\,mag (goal $U$\,$\ge$\,18\,mag), a S/N\,$=$\,20 at 313\,nm for a 0.007\,nm wavelength pixel. For different pixel sizes, the S/N ratio shall scale accordingly. The S/R value accounts the full light path efficiency (from atmosphere to detector), considering a telescope Cassegrain flat throughput of 75\% (ESO ETC database). Sky conditions are specified the footnote in Section \ref{sec:SNR}.
\end{itemize}

Considering these TLRs, we developed a baseline design for CUBES. In what follows, we first give a brief overview of the full CUBES system design, then details on each sub-system.

\subsection{CUBES system overview}
\label{sec:gen}
\begin{figure*}[h!]
\centering
  \includegraphics[width=0.95\textwidth]{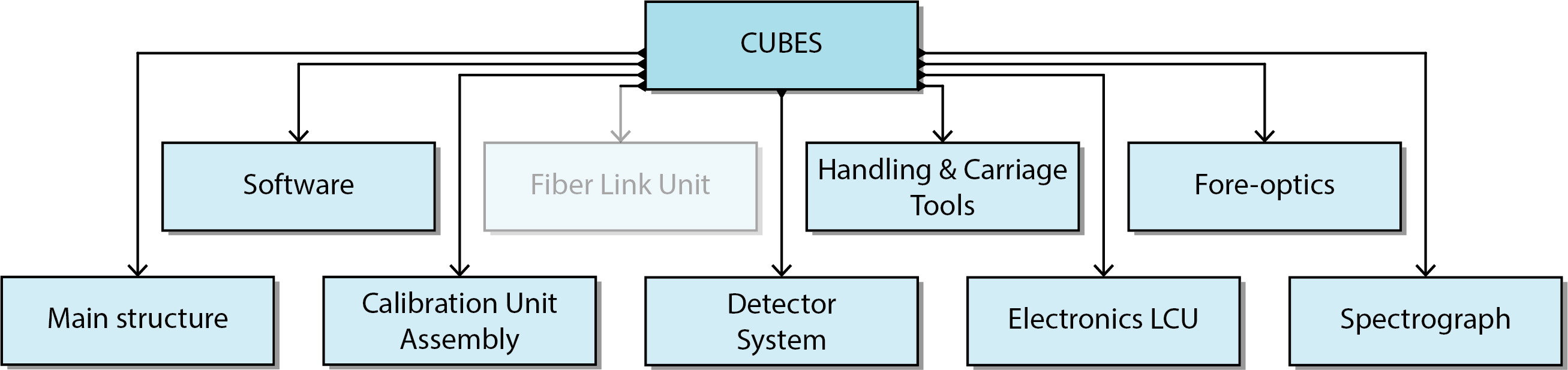}
\caption{Sub-system breakdown of the CUBES instrument. Fiber Link Unit is shaded since it is optional and has to be confirmed in Phase-B.}
\label{fig:bdd}
%\end{figure*}
\vspace{0.4in}
%\begin{figure*}[h!]
\centering
  \includegraphics[width=0.9\textwidth]{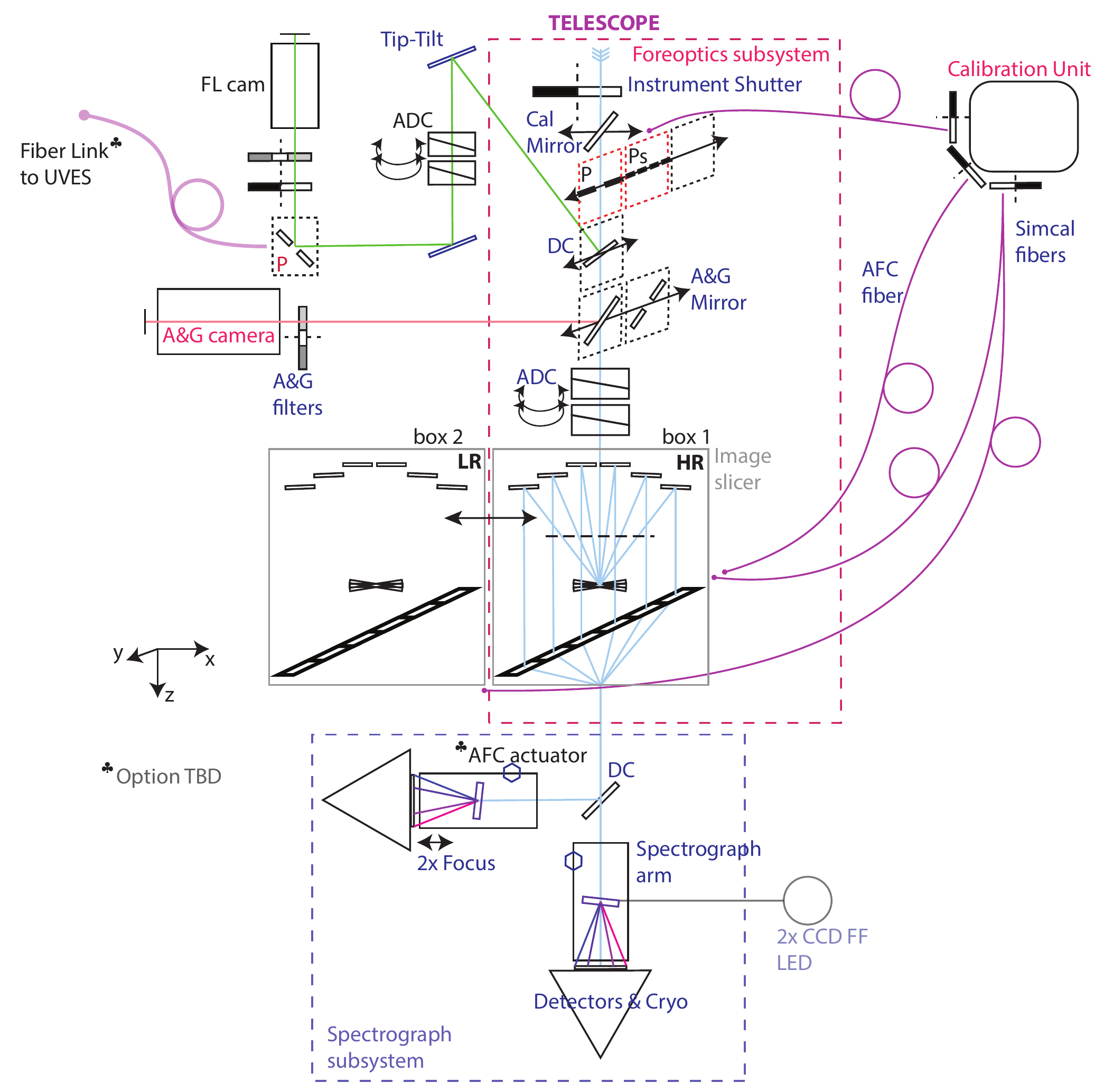}
\caption{Functional scheme of the CUBES system (in which the light path goes from top to the bottom). The following acronyms are used: DC - dichroic, P - pinhole, Ps - series of pinholes, A\&G - Acquisition and Guiding, AFC - Active Flexure Compensation system, FL - fiber link, ADC - Atmospheric Dispersion Corrector. Optional modules (TBD in Phase-B) are marked with a trefoil sign.}
\label{fig:scheme}
\end{figure*}

We organized the instrument into sub-systems and elements in a product-tree fashion, as shown in Fig.~\ref{fig:bdd}. The nine core units are independent in terms of the breakdown of work packages.

From analysis of the TLRs and taking into account the assembled science cases, we identified all the functions that CUBES should implement. The functional diagram of the instrument is shown in Fig.~\ref{fig:scheme},
%Functions are represented via icons and geometries following 
in which the light path goes from the telescope Cassegrain interface (top) to the detector (bottom). The baseline design of CUBES consists of the following components:

\begin{itemize}
    \item \textbf{Fore-optics:} provides a collimated beam for the Atmospheric Dispersion Corrector (ADC) and the magnifying optics to feed the image slicer.
    \item \textbf{Acquisition and Guiding (A\&G):} a separate camera fed by a fold mirror close to the telescope focus.
    \item \textbf{Image slicers:} two interchangeable slicers to enable different spectral resolutions: a high-resolution (HR) slicer to reformat the rectangular 1.5\,$\times$\,10\,arcsec Field of View (FoV) to generate the spectrograph entrance slit, and a second low-resolution (LR) slicer with a FoV of 6\,$\times$\,10\,arcsec.
    \item \textbf{Spectrograph:} includes a dichroic beamsplitter to spectrally separate the full bandwidth into two arms, then for each arm:
    \begin{itemize}
        \item[$\diamond$] two fold mirrors;
        \item[$\diamond$] a collimating lens;
        \item[$\diamond$] a transmission grating for the necessary spectral dispersion;
        \item[$\diamond$] a camera to focus the spectrum on the detectors.
    \end{itemize}
    \item \textbf{Detectors:} two CCDs (one for each arm) mounted in separate cryostats.
\end{itemize}

The light path is as follows. After the telescope focus and an instrument shutter, there is a calibration mirror, which is a moving element that can inject the light from the calibration unit. Next, a pin-hole mask is included for alignment purposes and to check focus during operations. An insertable dichroic follows, to relay longer-wavelength light to the FiberLink (FL) unit. Provision of the FL enables the instrument to operate in two modes: either using CUBES on its own or in combination with the FL for simultaneous observations with UVES at longer visible wavelengths.

Next, a double-position mirror slide can pass light to the A\&G system for the whole FoV or a peripheral part by moving to a holed position. This provides initial acquisition of the science field and can then be used for secondary guiding if required. Two counter-rotating prisms are used to then correct for atmospheric dispersion before relaying light to the slicer unit.

The Image Slicer unit reformats the FoV to generate narrower entrance slits for the spectrograph. This allows us to achieve the proposed resolution without the need for adaptive optics (which is not feasible in the near UV at present) and reduces the size of the optical components of the instrument. Some of the science cases of CUBES require excellent throughput but spectral resolution is less critical (e.g. some of the cases developed for observations of comets \cite{ExA_Opitom}, young stellar objects \cite{ExA_Alcala} and massive stars \cite{ExA_Evans}). The introduction of a LR option therefore provides a powerful complement to the primary (HR) option. By increasing the amount of light from the source entering the effective slit, the LR option will deliver greater sensitivity. We incorporated the LR option in the optical design by including a second interchangeable unit with physically larger slices (thus also sampling a larger on-sky area). The two image slicers enable $R$\,$\sim$\,24000 and $\sim$\,7000 for the HR and LR modes, respectively.

After the pseudo-slit the photons are split by another dichroic mirror, feeding the two arms of the spectrograph that finally collimate the beams, provide the dispersing function using transmission gratings, and use cameras to focus the spectra on the detectors (see more details in Sect.~\ref{sec:spe}).

An Active Flexure Compensation system (AFC) was considered in the design, in case analysis in the next phases shows that the required resolving power under a varying gravity vector cannot be achieved. This is not included in the baseline design of the instrument, but could be added in the next phase if needed. 

To identify which instrument layout best satisfies the requirements, dedicated simulation tools developed for CUBES \cite{Ref-Genoni_2021}, were used to investigate different instrument configurations (number of arms, detector type, presence of ADC and guiding systems, etc.).

The CCDs provisionally selected in the design have 9k\,$\times$\,9k 10\,$\mu$m pixels.
By setting the mean geometrical projection of the spectral resolution element to $\approx$\,25\,$\mu$m, i.e. 2.5\,pixels, allowed the wavelength range of the spectrograph design to be maximized to 300\,-\,405\,nm.

\subsection{Optical Design}
\label{Optical_Design}

\begin{figure*}[h!]
\centering
  \includegraphics[width=0.8\textwidth]{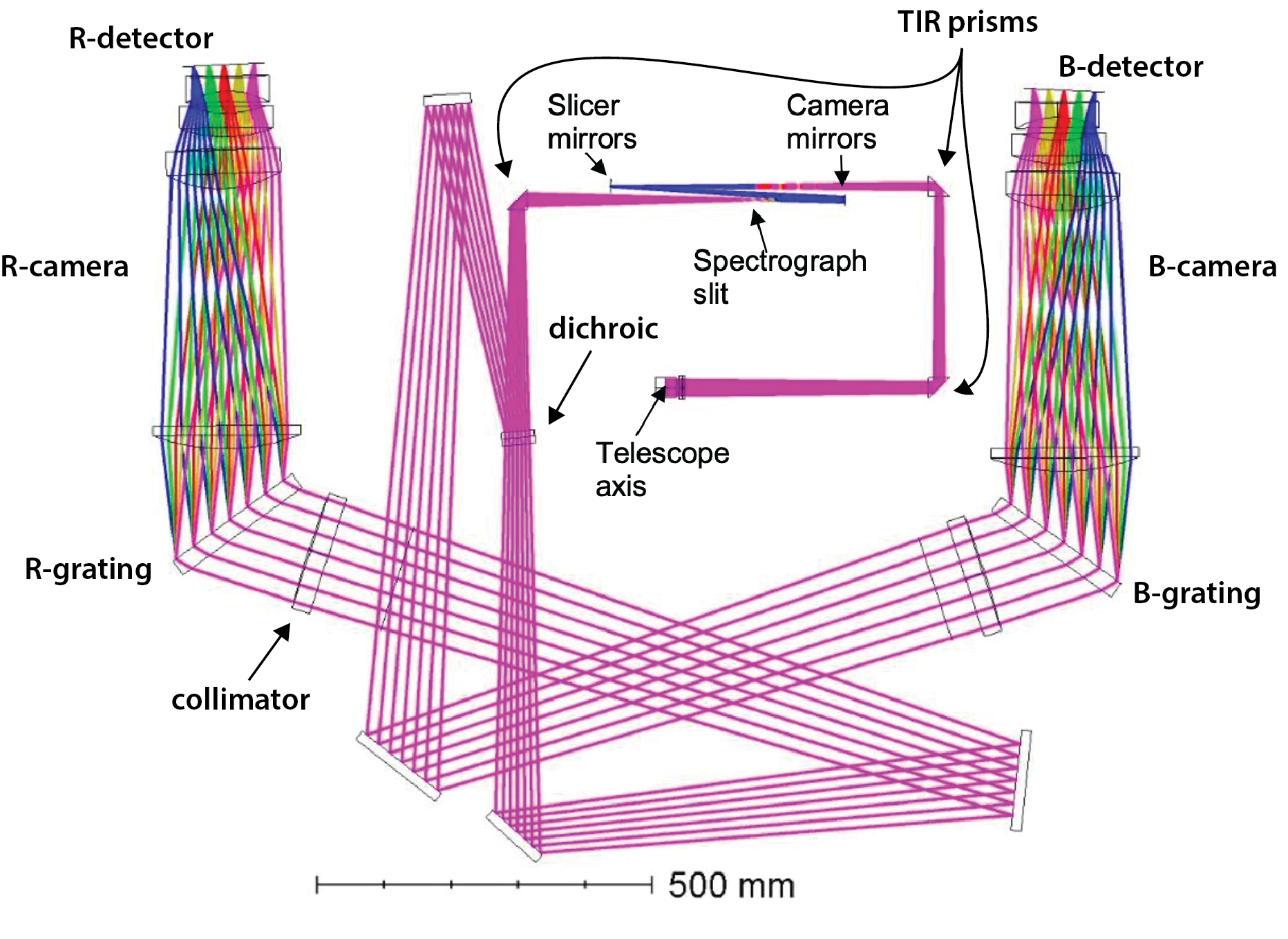}
\caption{Optical end-to-end model of the CUBES instrument.}
\label{fig:opt_e2e}
\end{figure*}

The end-to-end optical scheme of the instrument is presented in Fig.~\ref{fig:opt_e2e}.

\subsubsection{Fore-optics}
The fore-optics were designed and optimized against these criteria:

\begin{enumerate}
\item transfer and magnify a FoV of 6\,×\,10\,arcsec from the telescope focus to the Image Slicer at a scale of 0.5\,arcsec/mm;
\item correct for atmospheric dispersion over a wavelength range of 300\,--\,400\,nm for zenith angles of 0-60$^\circ$;
\item maintain a stable exit image and pupil at a wavelength of 346\,nm (chosen for calibration purposes using a Hg light source);
\item position the exit pupil to be 1160\,mm before the slicer image plane.
\end{enumerate}

Nominal environment values of T\,$=$\,10\,$^\circ$C and P\,$=$\,750 mbar were used in the model optimization. 

The full optical train from the telescope focus to the image slicer focal plane is shown in Fig.~\ref{fig:opt_for}. The first element is the A\&G pick-off mirror with a central hole for the science field to pass through to the spectrographs. A collimator doublet provides the parallel beam for the double-double prism type ADC, consisting of two fused silica/CaF2 cemented prism pairs. The input and output faces of the prism pairs are tilted at an angle of 0.90$^\circ$ with respect to the rotation axis while the angle of the common surface is at 25.22$^\circ$ to the rotation axis. The prism pairs are 30\,mm in diameter. The camera lens provides the focus at the input of the image slicer. All foldings of the science beam are obtained by total internal reflection prisms to maximize the throughput.

\subsubsection{Acquisition and guiding (A\&G) system}
Initial acquisition of the target will be achieved using the pointing capabilities of the VLT (accurate to 3\,arcsec rms). The A\&G unit will then provide more accurate positioning to account for offsets between the reference frame of the telescope from the FoV of the instrument.

\begin{figure*}[h!]
\centering
  \includegraphics[width=0.5\textwidth]{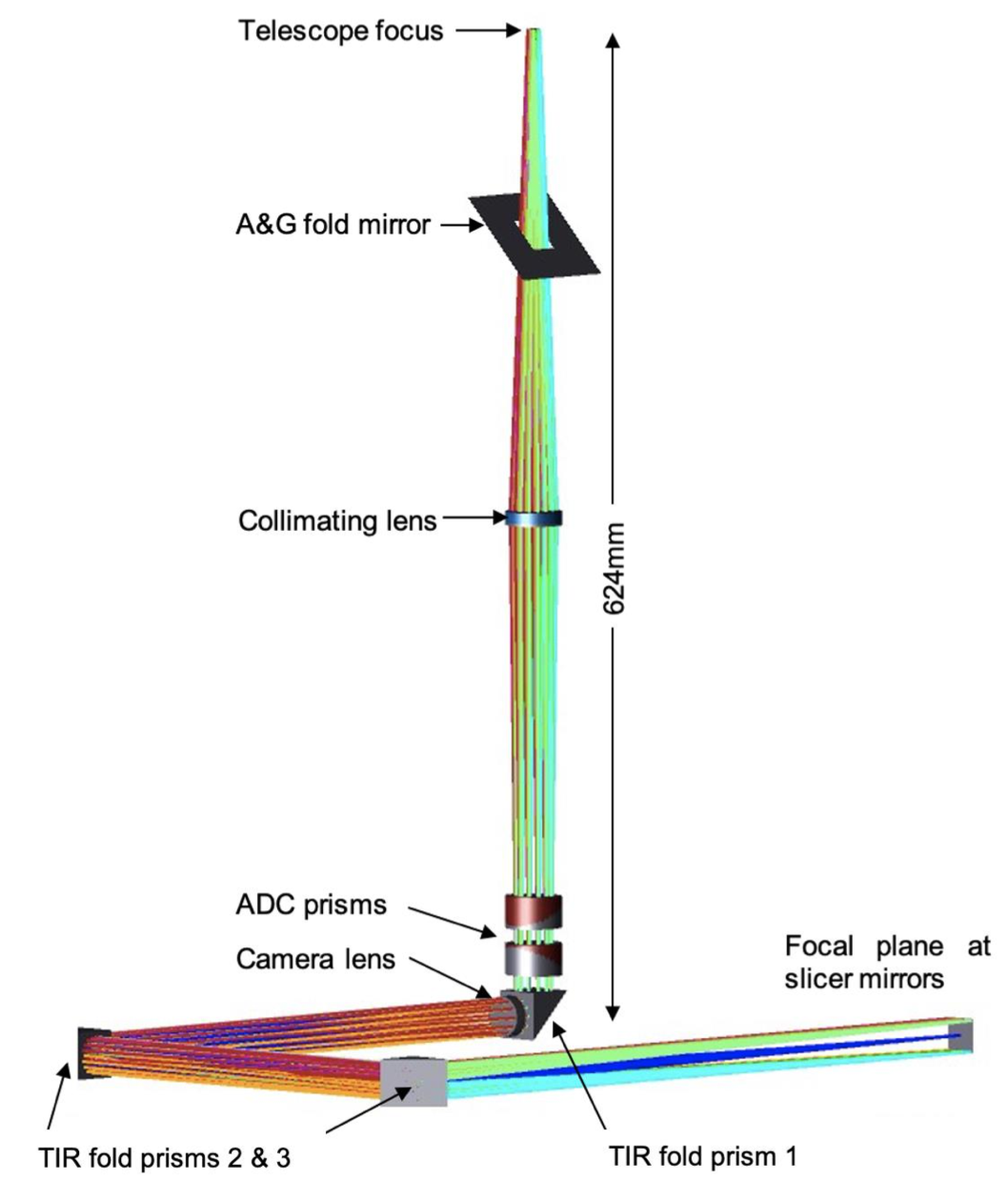}
\caption{3D optical scheme of the fore-optics sub-system of the CUBES instrument.}
\label{fig:opt_for}
\end{figure*}

This sub-system allows selection of the spectroscopic target from a 90\,$\times$\,90\,arcsec FoV by means of a mirror (45$^\circ$ fold with a central hole) placed off-center in the main optical train, to view the full FoV (see Figs~\ref{fig:scheme} and \ref{fig:opt_for}, and Sect.~\ref{sec:gen}).
The distance of the object from the field center (aligned with the Image Slicer entrance aperture) will be calculated and offsets sent to the telescope by the Telescope Control Software (TCS).

The user will then be asked to confirm a guide star whose position is periodically measured during integration and, if necessary, offset commands will be sent automatically to the telescope. When the science CCDs are integrating the A\&G mirror will be parked in its observing position that provides a hole to allow the central field to be transmitted to the spectrograph.

The A\&G system will use CCD cameras, selected from commercial-off-the-shelf components, based on the Gigabit Ethernet protocol that comply with the VLT standards for new instrumentation.

\subsubsection{Image slicers}
\begin{figure*}[h!]
\centering
  \includegraphics[width=0.8\textwidth]{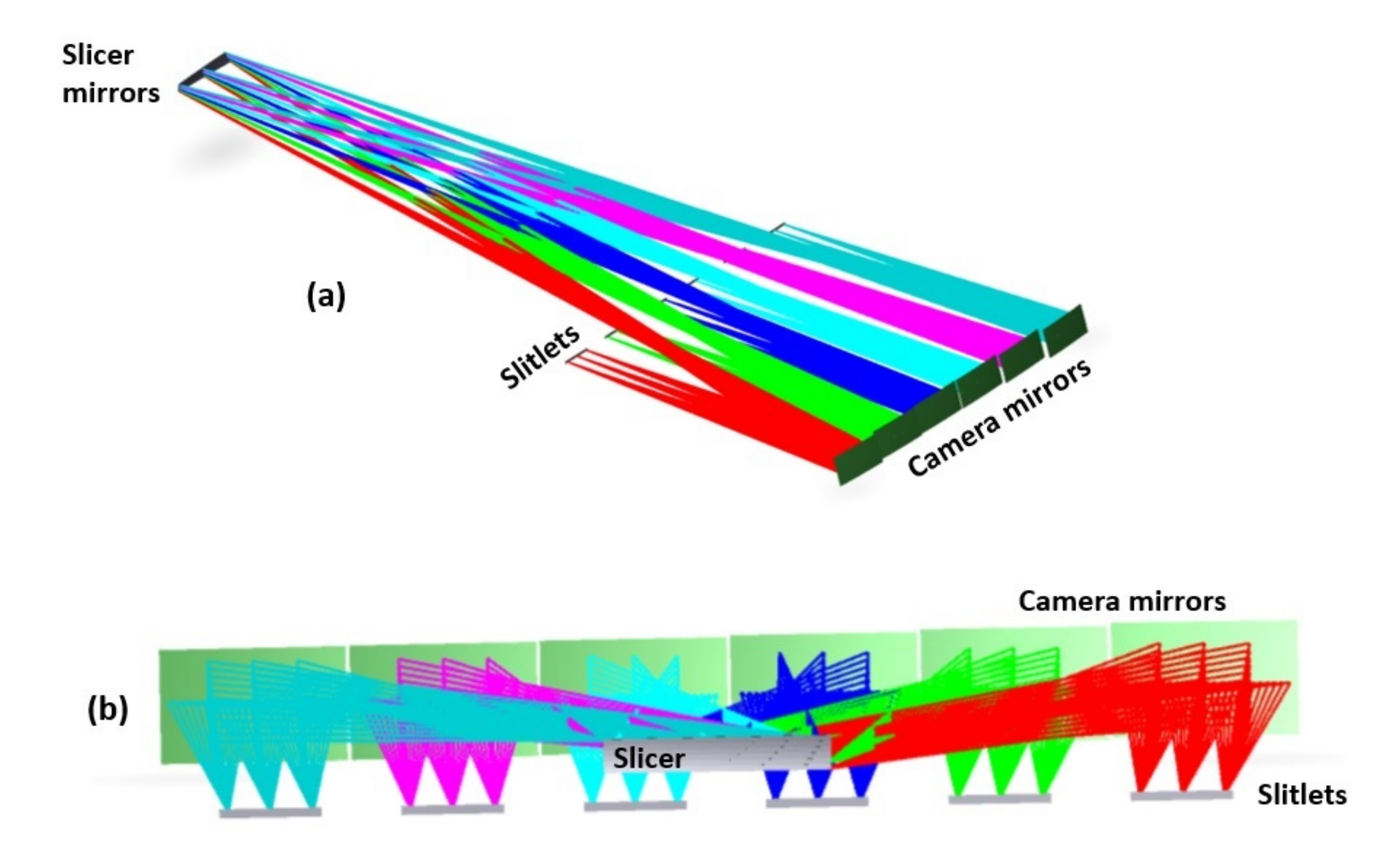}
\caption{(a) Optical design of the high-resolution (HR) mode image slicer. For both image slicers, the layout is composed of two arrays of six spherical mirrors each, the slicer mirror array and the camera mirror array. (b) The images of the six slices of the field of view generated by the image slicer, slitlets, compose the spectrograph entrance slit. The distribution of the slitlets, in order to define the slit shape and dimension, is enabled by the control of the orientation of the camera mirrors (tilt with respect to the X and Y axes). Further details about the image slicers design can be found in \cite{Ref-Calcines_2021}}
\label{fig:image slicer layout}
\end{figure*}

At the magnified telescope focal plane produced by the fore-optics, an image slicer decomposes the image of the FoV into six slices, which are re-organized composing the spectrograph entrance slit. Two image slicers are proposed to provide the HR ($R$\,$\sim$\,24000) and LR ($R$\,$\sim$\,7000) modes. Their layouts are conceptually identical. Both image slicers are a reflective sub-system composed of two arrays of six spherical mirrors each, called slicer mirrors and camera mirrors (Fig.~\ref{fig:image slicer layout} (a)). The slicer mirrors are thin rectangular powered mirrors, each with a different orientation (tilt angle with respect to the X and Y axes) to reflect a slice of the FoV in a different direction towards a camera mirror. The camera mirrors are rectangular spherical mirrors, whose power is used to focus the beams. Thus, each camera mirror generates an image of its corresponding slice of the field of view, called slitlet, at its focal length (Fig.~\ref{fig:image slicer layout} (b)). Each camera mirror is provided with a tilt angle with respect to the X and Y axes, which enables control of the distribution of the slitlets. These six slitlets, separated by five gaps between them, compose the spectrograph entrance slit. The capability of controlling their orientation enables their alignment and the overlapping of the six exit pupils. The image slicer design is highly efficient, using the minimum number of surfaces possible (two). These will be made of Zerodur with an optimized dielectric coating to achieve a reflectivity of 95\% per surface, with a goal of 97\%. Each slicer will be stand-alone and contained within a housing. The two housings will be implemented on a motorized linear stage, which will allow users to select the resolution. For further details of the slicer design and layout see \cite{Ref-Calcines_2021}.

\subsubsection{Spectrograph}
\label{sec:spe}
The current baseline of the spectrograph sub-system is for two arms with common optics before a dichroic, which splits the light by reflecting the Blue-Arm passband (300\,--\,352.3\,nm) and transmitting the Red-Arm passband (346.3\,--\,405\,nm).

The input slit consists of six slitlets, one for each slice in the image slicer (0.25\,$\times$\,10\,arcsec on the sky for HR mode, and 1\,$\times$\,10\,arcsec for the LR mode) and three fiber sources (two for simultaneous wavelength calibration and one for the AFC).
The layout of the two arms after the dichroic is similar but the individual components and separations differ so as to achieve the required dispersion, magnification and image quality for the two passbands using only fused silica.

CUBES uses first-order gratings produced by microlithographic techniques with an average efficiency of 85\% over the whole wavelength range (see \cite{Ref-Gratings_2021} for further details).

The spectrograph camera is composed of four lenses, the last of which acts as a window to the detector cryostat. Of the eight optical surfaces in these components, four are spherical and four have modest conic constants (between $-$1 and 0.560). To correct for thermal changes in the focus of the spectrographs the first three lenses in each camera will be moved as a single unit. The science CCDs are tilted by 3.80$^\circ$ and 2.45$^\circ$ for the blue and red arms, respectively. If needed, the AFC will be implemented  by moving the collimator lenses in X and Y.

\subsection{Mechanical Design}
CUBES requires a beam diameter of 160\,mm. The instrument envelope is therefore fairly large compared to other Cassegrain instruments (e.g. X-Shooter \cite{Ref-XSH} with a 100\,mm beam diameter). Scaling classical instrument designs to the required size of CUBES would then exceed the mass limit of 2500\,kg for Cassegrain instruments of the VLT Unit Telescopes (UTs). Therefore, the mechanical design of CUBES adopts light-weight construction principles and makes use of modern composite materials.
The optical layout was optimized such that all optical elements of the spectrograph from slit to detector lie in a single plane, so all the spectrograph optics can be mounted on a single optical bench of size 1.3\,×\,1.7\,m. This is arguably the most stable configuration since the dispersion direction of CUBES is parallel to the stiff surface plane of the optical bench.
Another focus of the mechanical design is to minimize the effects of gravitational bending
of the instrument.

In the current design, the CUBES mechanical structure is divided into three main components:
\begin{enumerate}
    \item \textbf{Telescope adapter:} provides a stiff connection between the Cassegrain telescope flange and the optical bench assembly of CUBES;
    \item \textbf{Optical bench assembly:} provides a stable platform for the spectrograph optics as well as for the fore-optics;
    \item \textbf{Support frame assembly:} provides support for auxiliary equipment of CUBES such as the electronic racks, the calibration unit and vacuum equipment. This frame is detached from the optical bench assembly to mitigate flexure contribution from auxiliary equipment.  
\end{enumerate}

As for the materials, we plan to use steel for the telescope adapter and the support frame, and a Carbon-fiber-reinforced polymer (CFRP) for the optical bench. We plan to make the opto-mechanics of aluminum alloys (e.g., AlSi40), to improve the specific stiffness and lower the mismatch in coefficient of thermal expansion between the opto-mechanical parts and the CFRP bench (in case thermal analysis in the next phases shows this is an issue).

Initial finite element method (FEM) analysis shows that the global structure (bench + telescope adapter) has no need for an AFC. However, it requires further investigation if this assumption holds when taking gravitational flexure of individual components (in particular of the heavy CCD heads) and its interplay with optics into account.

\begin{figure*}[h!]
\centering
  \includegraphics[width=0.5\textwidth]{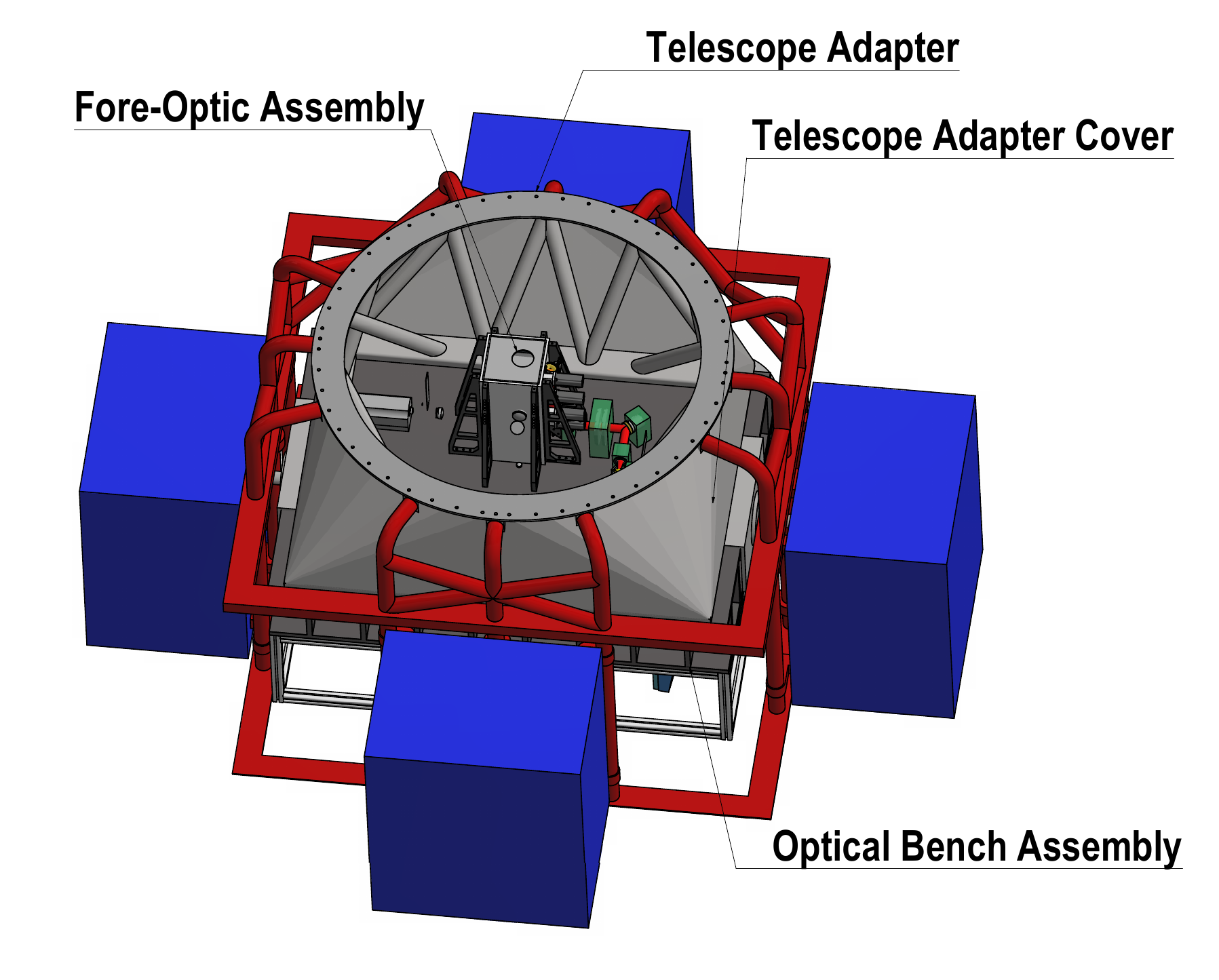}
  \includegraphics[width=0.65\textwidth]{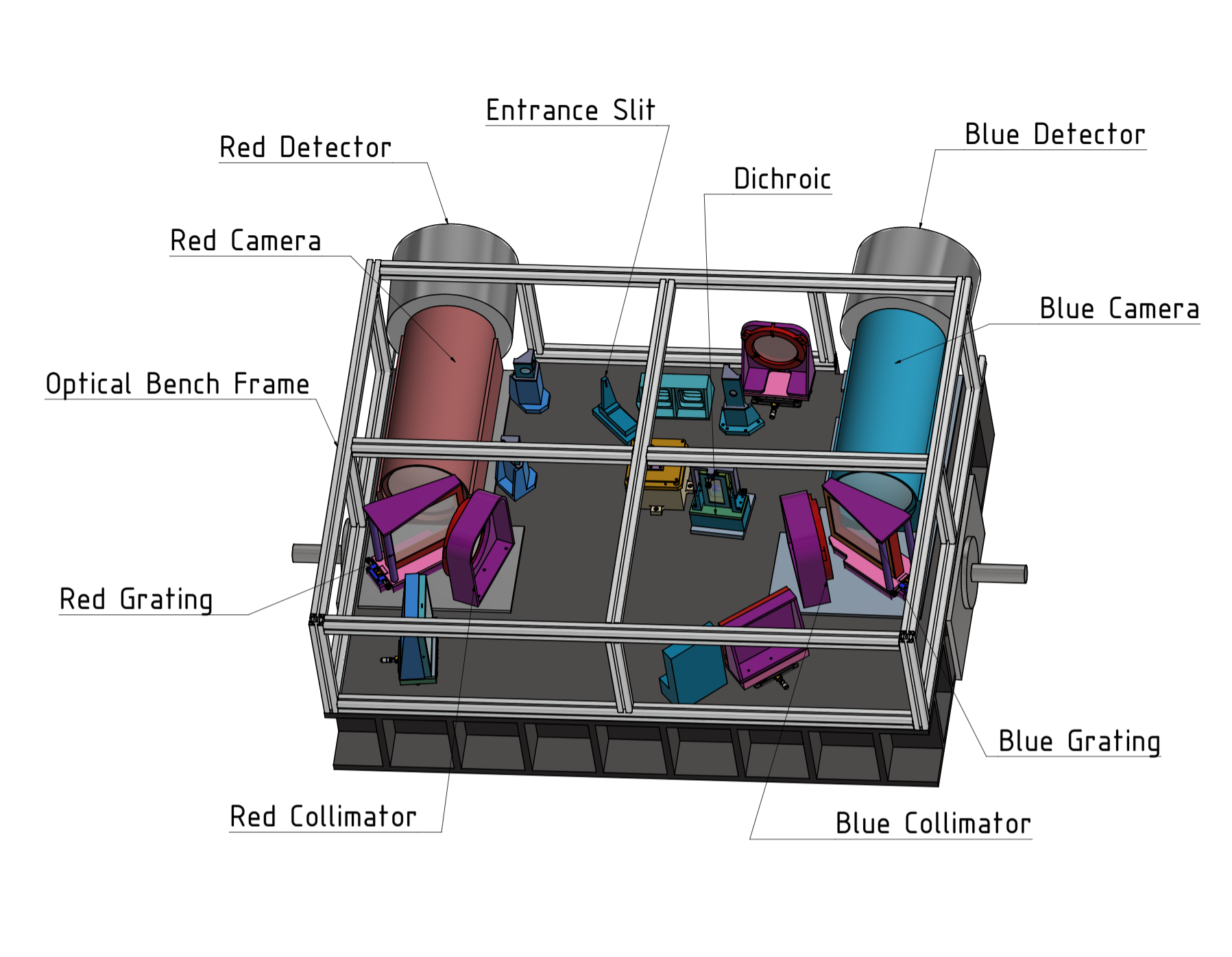}
\caption{Mechanical concept for CUBES: The general layout of the main mechanical components is shown in the upper image. For reference, the Telescope Adapter has a diameter of about 1.5\,m. The Fore-Optics Assembly is located near the center of the bench, while the spectrograph opto-mechanics are `hanging' on the bottom side of the same bench (lower image).}
\label{fig:mech}
\end{figure*}

\subsection{FiberLink (FL) Unit}
The FL will provide the option of simultaneous observations with UVES, a two-arm, high-resolution echelle spectrograph located at the Nasmyth B focus of VLT-UT2. A similar concept, HiRISE \cite{Ref-HiRISE}, is currently under development that combines SPHERE and CRIRES$+$ via optical fibers. 

\begin{figure*}[!t]
\centering
  \includegraphics[width=0.6\textwidth]{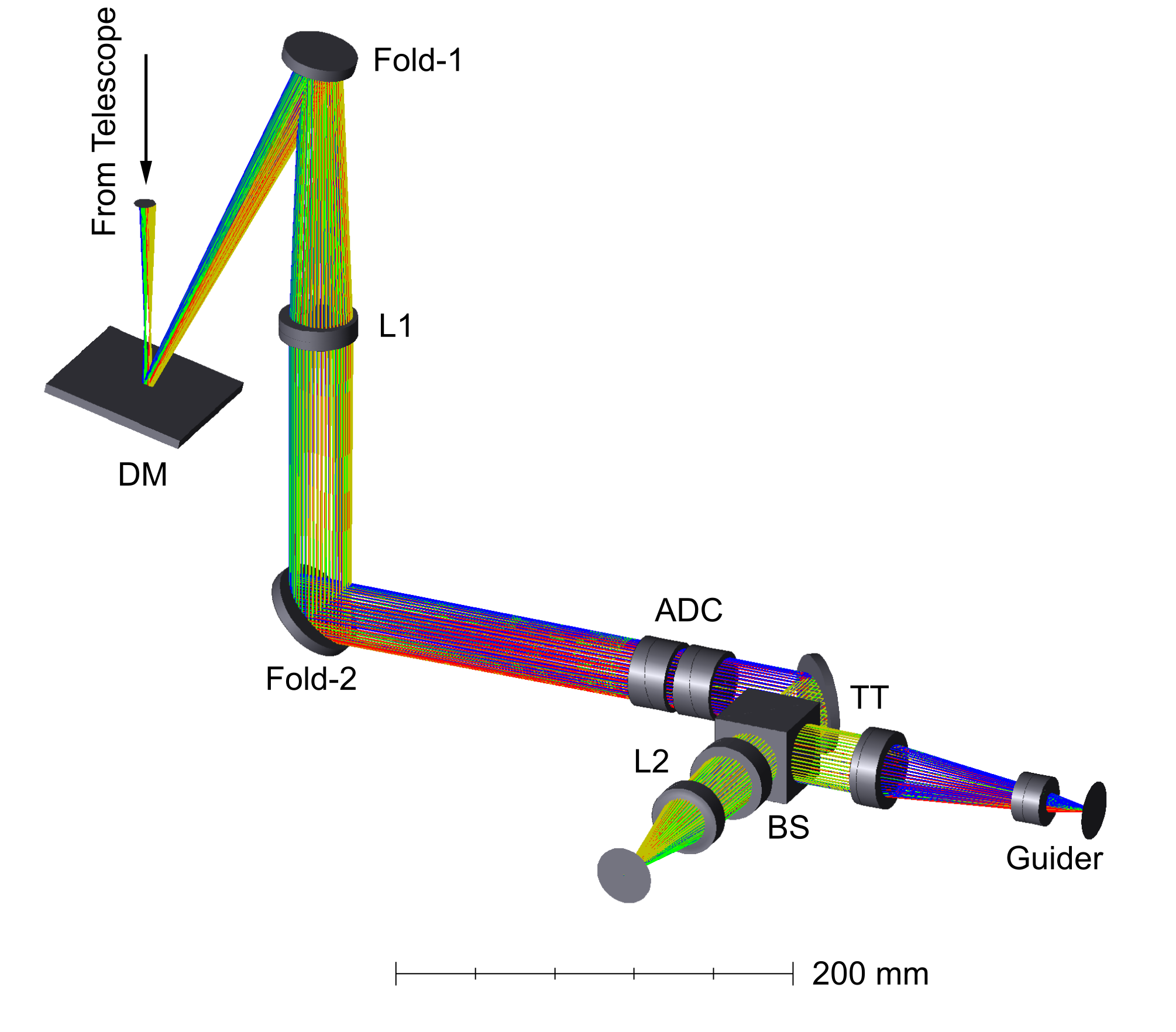}
\caption{FiberLink FE module: A doublet (L1), placed behind Fold-1 and dichroic mirror (DM)} collimates the beam, sending it through an ADC while passing a second fold mirror (Fold-2). A fast tip-tilt (TT) mirror, located in the pupil of L1, folds the optical path towards the camera (L2) which images the field at F/3 telecentrically onto the fiber tips.
\label{fig:FL1}
\end{figure*}

To enable simultaneous observations with UVES, the CUBES fore-optics will be equipped with a front-end (FE) module that feeds up to seven optical fibers with stellar, sky-background, and calibration light. The optical fibers have a 120\,$\mu$m core that subtends a 1-arcsec aperture on the sky. The light is picked-off using a dichroic mirror (DM) which can be moved into the telescope beam. The DM diverts light redder than 420\,nm into the FE module. The FE module will contain subsystems for field stabilization (tip/tilt), an ADC, and fiber guiding (FG). The relay optics of the FE module convert the native F/13.41 telescope beam into an F/3 beam to minimize loss due to focal ratio degradation (FRD) \cite{Ref-FL}. A schematic diagram that briefly outlines the conceptual design of the FE optics is shown in Fig.~\ref{fig:FL1}.

A 40\,m long fiber cable installed at the UT2 telescope will transmit light from the Cassegrain flange to UVES. The UVES pre-slit optics nominally relay the UT2 Nasmyth beam into its blue (300--500\,nm) and red (420--1100\,nm) arms. The injected light from the CUBES fibers will be directed into the red arm only through a fold mirror mounted on the back of the dichroic mirror, which usually splits the light into the two wavelength bands; this is a
similar approach to that used for the FLAMES fiber-feed to UVES, which also only uses the red arm \cite{Ref-FLAMES}.

\subsection{Calibration Unit}
The calibration sub-system provides the light sources necessary to register frames for flat fielding, wavelength calibration, the option of simultaneous wavelength calibration, alignment and the AFC option if required. This unit will be mounted on one of the side cabinets and will feed light into the spectrograph by means of optical fibers in two ways: on the fore-optics by using a folding mirror (for daytime calibrations), and at the slit level at the sides of the science spectra (for simultaneous, night time calibrations).

%The calibration unit is in charge to provide light source in order to illuminate the spectrograph for recording flat field frames, wavelength calibration frames as well as to enable the simultaneous calibration, AFC verification, linking CUBES with UVES and provide sources for alignment purposes. The calibration plan foresees both daytime and night time calibrations.

\begin{figure*}[h!]
\centering
  \includegraphics[width=0.7\textwidth]{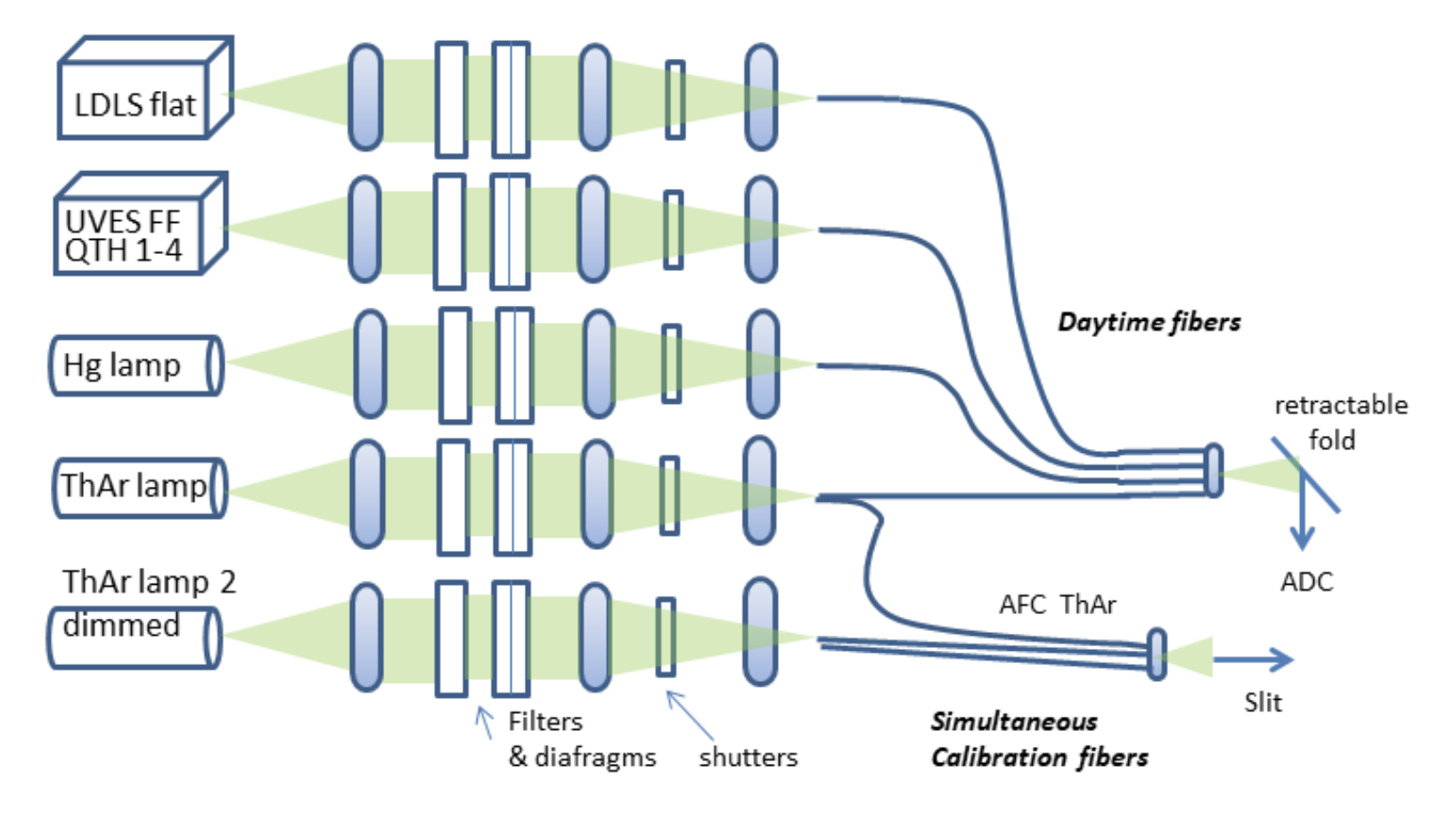}
\caption{Schematic concept of the calibration unit.}
\label{fig:cal1}
\end{figure*}

\begin{figure*}[h!]
\centering
  \includegraphics[width=0.8\textwidth]{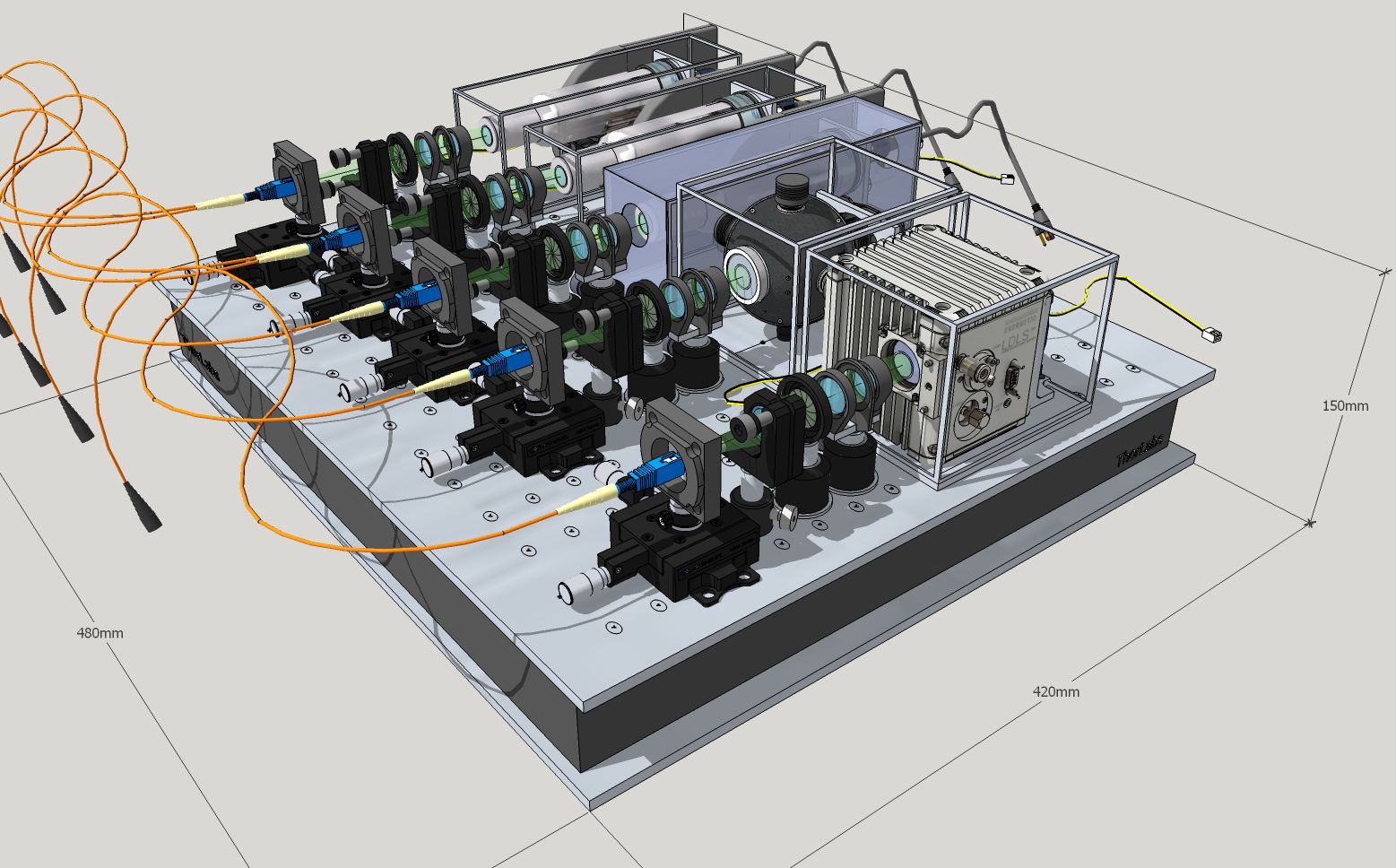}
\caption{3D mechanical concept for the lamp assembly of the calibration unit.}
\label{fig:cal2}
\end{figure*}

We will use two ThAr lamps (one dimmed for simultaneous calibration), one Hg lamp and one Laser Driven Light Source (LDLS) for a near UV flat and four QTH lamps for the UVES flat (TBC). In Fig.~\ref{fig:cal1} we show the schematic concept of the calibration unit showing the two optical-fiber bundles, one for the daytime calibration (including UVES) and alignment, and the other for wavelength calibration during an observation and to provide an offset measurement to a flexure compensation system if this is shown to be necessary to achieve the required resolving power. In Fig.~\ref{fig:cal2} we show the concept for the mounting of the lamps and optics in the calibration box.

\subsection{Detector and Electronics}
The baseline detectors are two Teledyne-e2v CCD290-99 (9k\,$\times$\,9k) arrays \cite{Ref-CCD,Ref-CCD-data}, one for each of the blue and red arms. These are standard silicon, backside illuminated devices, that will use an anti-reflection coating now in development at Teledyne-e2v (with optimized performance over the 300\,--\,400\,nm range). The detector control electronics will be the New Generation Controller II (NGCII) that is currently under development at ESO \cite{Ref-ESODET}. 

The science data on the CUBES arrays will only occupy a small percentage of the available rows, with a 14\,$\times$\,92\,mm footprint of the spectra on the detectors. This equates to the full width of the device but only 1400 rows vertically. This spectral science data will be projected onto the lower half of the detectors.  A region of the upper half of the devices could be used to support the AFC system if it is required, with a windowed part of the array read out at a faster frame rate than the science data. As such, independent control of the upper and lower halves of the CCDs is required.

The dark current (DC) of the detectors is a critical performance requirement. The DC level specified by Teledyne-e2v for the CCD290-99 arrays is 3\,e$^-$/pix/hr at 173\,K. The target
value for CUBES is $<$\,0.5\,e$^-$/pix/hr (to enable, e.g., background-limited performance to $i$\,$\ge$ 20\,mag for high-redshift galaxies), so the arrays will need to be cooled to a lower temperature of $\le$\,165\,K.

The baseline cryostat design is a liquid nitrogen bath that is similar to others in use at ESO. Individual baths will be fitted to each detector cryostat, adopting a similar layout to that used for the X-Shooter cryostats. The evacuated volume of each cryostat is $\sim$\,3\,l, so the 1\,-\,10\,l standard ESO vacuum system will be used. The cryostat control electronics will control pumping and venting, detector thermal control, cold finger thermal control (used for warm-up and coarse temperature control), warm-up control, liquid-nitrogen level monitoring, monitoring of the cryostat states and logging data, and alarms and safety-critical interlocks.

\section{Software}
\subsection{Control Software}
The CUBES instrument control software is based on the ELT Instrument Control Software Framework (IFW), a new toolkit developed by ESO aimed to help instrument developers implementing the control software for the ELT instruments. The IFW reuses the proven architectural and design patterns from the VLT control software framework but is implemented using the new technologies defined by the ELT development standards, such as the use of Beckhoff Programmable Logic Controllers (PLCs) with OPC-UA as the communication protocol, C++ 17 and Python as programming languages, ZMQ as middleware and the use of Nomad and Consul for the management of the life cycle of the software components.

\begin{figure*}[h!]
\centering
  \includegraphics[width=0.8\textwidth]{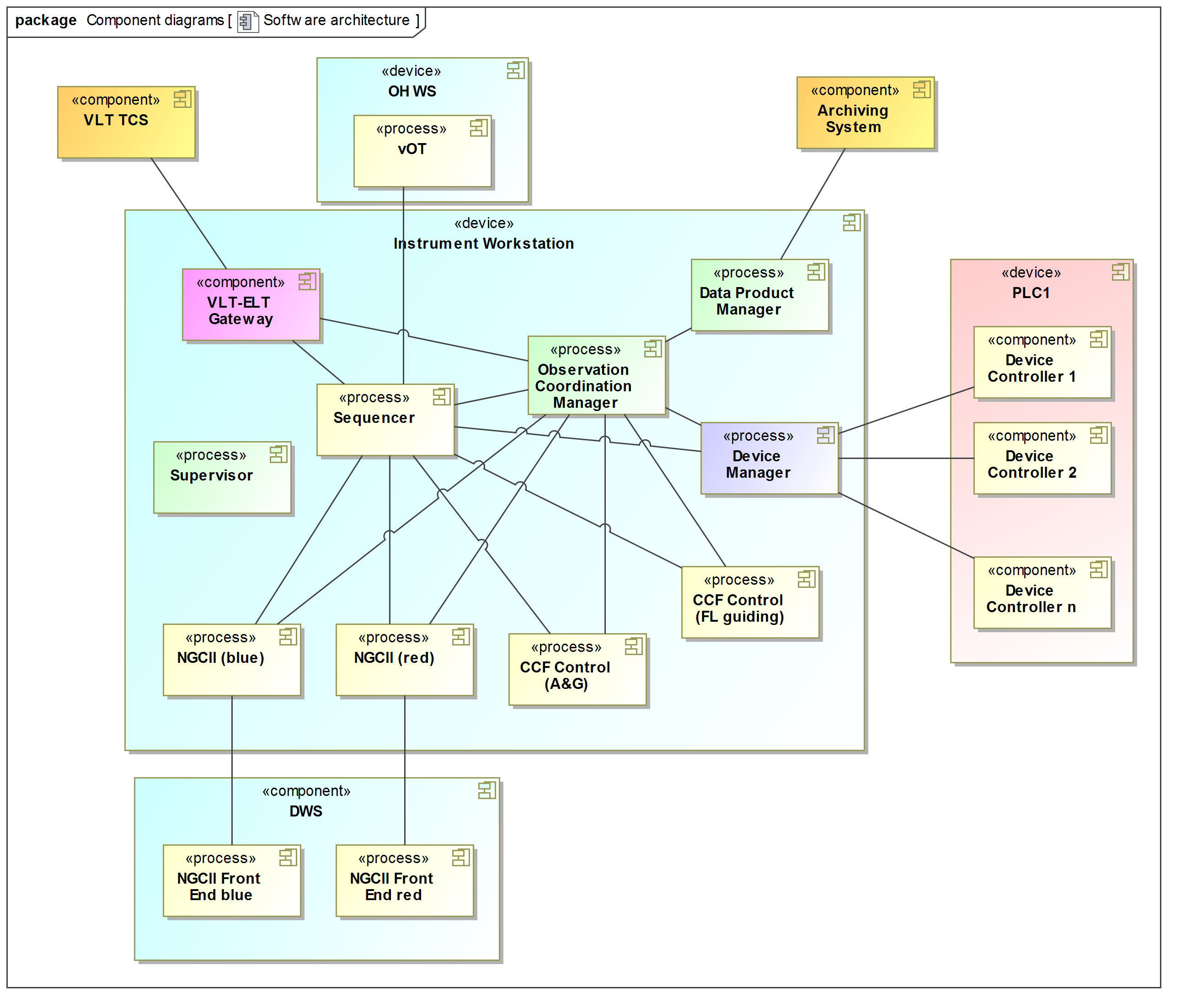}
\caption{CUBES software architecture. For a complete description of the elements refer to the section \ref{sec:SW_arch}.}
\label{fig:sw_arch}
\end{figure*}

The IFW is composed by the following ELT standard instrument software subsystems:
\begin{itemize}
    \item {\bf Function Control System (FCS):} controls and monitors the instrument hardware functions and sensors, except the detectors. 
    \item {\bf Detector Control System (DCS):} carries out all the tasks to control the detector subsystems.
    \item {\bf Observation Coordination System (OCS):} the highest layer of the control software. It consists of a Supervisor, responsible for the overall monitoring and supervision of the instrument subsystems, an Observation Coordination Manager, responsible for coordinating the scientific exposure and a Data Product Manager, responsible for collecting and archiving the results of the scientific exposure in the final FITS file and for sending it to the Online Archive System.
    \item {\bf Maintenance Software (MS):} used for instrument configuration and maintenance procedures.
\end{itemize}

As CUBES will be installed on one of the VLT UTs, it will need to interface its ELT IFW-based control system with the VLT physical environment. To this end, a dedicated intermediary or gateway, the VLT-ELT Software Gateway component developed by ESO, will translate protocols and interfaces between the ELT and VLT environments, allowing in particular the instrument communication with the VLT TCS. For simultaneous observations with CUBES and UVES, the VLT-ELT Software Gateway will provide the synchronization between the CUBES and UVES on-line databases (OLDBs).

\subsection{CUBES software architecture}
\label{sec:SW_arch}
The CUBES software architecture is shown in Fig.~\ref{fig:sw_arch}.
The Observation Blocks (fundamental scheduling units of VLT science operations) are transferred from the (visitor mode) Observing Tool (vOT) hosted on the Observation Handling Workstation (OH WS) to the Sequencer which reads the contents of the Observations Blocks and executes one by one the templates specified in there. Each template consists in a sequence of commands to be sent to the various instrument subsystems (OCS, FCS, DCSs and, through the VLT-ELT Gateway, VLT Telescope Control System (TCS)). The subsystems then configure the corresponding hardware according to the settings that have been selected by the user. The Observation Coordination Manager (OCM) coordinates the data acquisition (data coming from the scientific detectors), triggers the creation of metadata from the other instrument subsystems and controls the data products creation.
At the end of the scientific exposure, the DCS generates the detector data and saves them in a FITS file. Then the Data Product Manager, controlled by the OCM, collects the CUBES data from the instruments subsystems, creates the final FITS file and transfers the file to the Archiving System. A Supervisor monitors and manages the lifecycle and states of the different subsystems. The two scientific detectors are controlled by one dedicated controller (New General detector Controller II, NGCII) and the technical cameras of the CUBES Acquisition \& Guiding (A\&G) System and Fiber Link (FL) Guiding System are controlled by the ESO standard Camera Control Framework (CCF).

\subsection{Data Reduction}
The Data Reduction Software (DRS) will extract science-grade spectra from the raw science and calibration frames produced by the instrument, removing the instrument signature.
The DRS will be written in ANSI C using the ESO Common Pipeline Library (CPL) and High-level Data Reduction Library (HDRL). It will be operated through the standard ESO interfaces (EsoRex, Gasgano, and ESO Reflex). An additional Python interface will be developed internally for better integration within the Python ecosystem, which is now a de-facto standard for the astronomical community.

\subsection{Simulation Tools}
An End-to-End (E2E) instrument simulator and an Exposure Time Calculator (ETC) have been developed to help define the current baseline design as well as in the scientific evaluation of the various observing modes.

The ETC is a tool to predict the global performances of the CUBES spectrograph considering the input instrumental parameters and the environmental conditions.
The ETC predicts the S/N ratio per pixel that is achievable for a given wavelength, exposure time, sky conditions, and magnitude ($U$, $V$) of a point source. It can also provide the exposure time required to obtain a given S/N ratio. The ETC is a Python 3.7.6 script that is called by a web application written in Java and JavaScript. 

The E2E simulator is a tool which aims to simulate the instrument behavior and the astronomical observations given the flux distribution of the scientific sources/targets (or calibration sources in the case of calibration frames) to the output raw-frame data produced by the CCD(s). 
Two different versions, targeted for different applications and users are under development:
\begin{itemize}
    \item Science version: for initial science evaluations and preliminary performances.
    \item Technical version: for more precise simulations to be exploited for DRS development and to aid in evaluating the instrument design and performances.
\end{itemize}

\section{Project Organization and Schedule}

\subsection{Consortium}
The current project builds on previous work from a joint Brazil-ESO Phase~A study of a near-UV spectrograph undertaken in 2012 \cite{Ref-Barbuy,Barb_2014,Ref-Bristow}. Unfortunately the project did not proceed further at the time, but the science case is now stronger than ever, as demonstrated by the range of papers presented in this Special Issue. 

Given the clear demand from the community for such a capability and the unrivalled ground-UV performance that it would provide for the Paranal Observatory, ESO issued a Call for Proposals for such an instrument in January 2020. In response to this call we assembled a new consortium to design and build CUBES, comprising institutes from six countries:

\begin{itemize}
    \item[$\circ$] INAF -- Istituto Nazionale di Astrofisica, \textit{Italy} (consortium lead).
    \item[$\circ$] UK\,ATC -- UK Astronomy Technology Centre (primary UK partner) and Durham University Centre for Advanced Instrumentation (secondary UK partner), \textit{United Kingdom}.
    \item[$\circ$] LSW -- Landessternwarte, Zentrum für Astronomie der Universität Heidelberg, \textit{Germany}.
    \item[$\circ$] NCAC -- Nicolaus Copernicus Astronomical Center of the Polish Academy of Sciences, \textit{Poland}.
    \item[$\circ$] IAG-USP -- Instituto de Astronomia, Geofísica e Ciências Atmosféricas (primary Brazilian partner) and LNA -- Laboratório Nacional de Astrofísica (secondary Brazilian partner), \textit{Brazil}.
    \item[$\circ$] AAO-MQ -- Australian Astronomical Optics, Macquarie University, \textit{Australia}.
\end{itemize}

\subsection{Schedule}
The Phase~A study commenced in June 2020, with a close-out review in June 2021.  The next phase of the project will be $\sim$2.5\,yrs for the detailed design, with two formal review milestones of the Preliminary and Final Design Reviews.  The project will then enter the Manufacturing, Assembly, Integration, Testing (MAIT) phase, with the next major milestone being the system Integration and Test Readiness Review (ITRR), that will then be closed by the Preliminary Acceptance Europe (PAE). CUBES will be then moved to Chile, re-assembled, mounted on the VLT and tested. The commissioning phase will be closed when the Provisional Acceptance Chile (PAC) is granted. Finally, CUBES will be offered to the community after PAC is granted and science verification is concluded. The MAIT phase through to science operations in the current plan is envisaged for 3\,yrs, meaning that it would be available to the ESO user community in 2028.

\section{Instrument performance}

\subsection{Throughput}
For an instrument working in the UV range at the limit of the atmospheric transmission, a high throughput is mandatory and needs special attention.
The aspects of this taken into consideration in the Phase A study are now briefly summarized.

The grating efficiency is one of the main factors on the total throughput of the instrument. A significant gain can be achieved by splitting the covered wavelength range into two channels, each with an optimized grating with efficiencies of $\sim$90\% over their respective wavelength ranges.

\begin{table}[h!]
\begin{center}
\caption{Adopted efficiencies in the two scenarios considered for the throughput analysis. `AR coating' (anti-reflection) refers to all transmissive surfaces: lenses, fold prisms, ADC prism pairs; `Cement' is transmission of the cemented surfaces (ADC prisms, collimator lens, camera lens doublets), `High refl.' refers to the high reflectivity coating of the two mirrors in each arm of the spectrograph and of the image slicer mirrors.}\label{tab:coat_perf}
\begin{tabular}{lcc}
\hline \hline
 & \textit{Optimum case} & \textit{Worst case} \\ \Xhline{1pt}
AR coating & 0.997 & 0.995 \\ \Xhline{0.2pt}
Cement & 0.995 & 0.99 \\ \Xhline{0.2pt}
\makecell[l]{High refl.\\mirror coating} & 0.98 & 0.97 \\ \Xhline{0.2pt}
Dichroic coating & 0.97 & 0.97 \\ \Xhline{0.2pt}
\makecell[l]{High refl.\\slicer mirror coating} & 0.97 & 0.95 \\ \Xhline{0.2pt}
Detector coating & \makecell{datasheet\\QE \cite{Ref-CCD-data}} & \makecell{datasheet\\QE \cite{Ref-CCD-data}\\ × 0.95} \\ \Xhline{0.2pt}
Slicer vignetting & 0.98 & 0.97 \\ \Xhline{0.2pt}
\makecell[l]{Grating\\(w/AR coating\\backside)} & \makecell{manufacturer's\\ simulation \cite{Ref-Gratings_2021}\\ × 0.95} & \makecell{manufacturer's\\ simulation \cite{Ref-Gratings_2021}\\ × 0.90} \\ \hline
\end{tabular}
\end{center}
\end{table}

\begin{figure*}[h!]
\centering
  \includegraphics[width=0.7\textwidth]{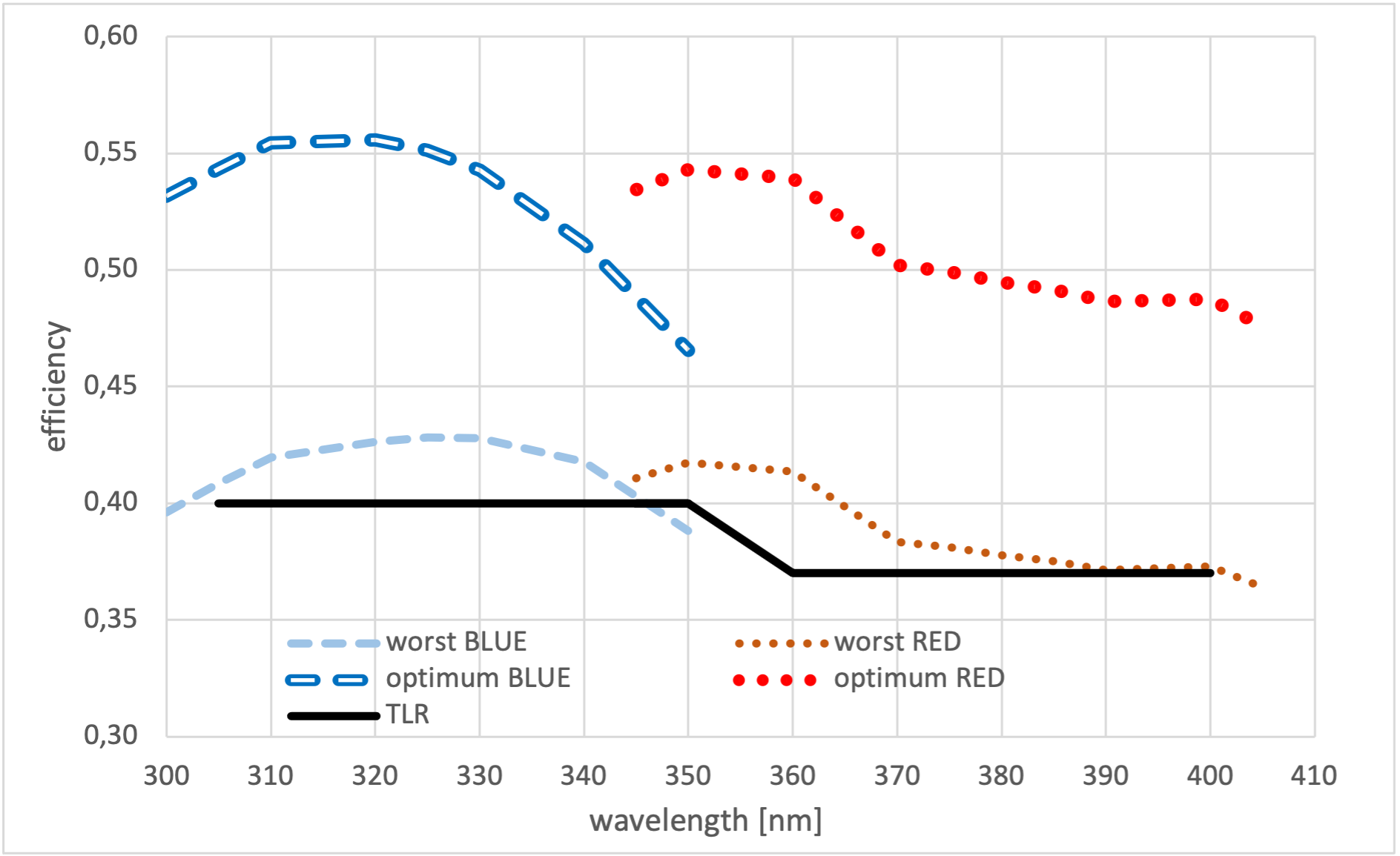}
\caption{Calculated detective quantum efficiency of the two arms of CUBES (BLUE and RED) for \textit{worst} and \textit{optimum} scenarios. The black line indicates the formal top-level efficiency requirement. The throughput was calculated from the telescope focus, including detector QE, IS vignetting, but excluding atmospheric effects and telescope losses.}
\label{fig:opt_eff}
\end{figure*}

The efficiency requirements also put restrictions on the selection of the materials of the optical elements (mainly Fused Silica with a few CaF$_2$ substrates) and infers the need to minimize the number of surfaces used. For this reason, special attention was also given to the selection of the anti-reflection (and mirror) coatings. Moreover, a detailed design was performed of the image slicer to ensure the highest efficiency.

During the conceptual design phase, first iterations with coating companies have been established to predict and simulate the throughput contributions of each optical element in the instrument (e.g. lenses, mirrors, gratings, and image slicer).

In many cases there were uncertainties regarding the achievable throughput values, so we estimated the end-to-end efficiencies for two scenarios. We therefore considered the max/min envelopes of the efficiencies by deriving \textit{optimum} and \textit{worst-case} values for the throughput of the optical elements (see Table \ref{tab:coat_perf} and Fig.~\ref{fig:opt_eff}).

As shown in Fig.~\ref{fig:opt_eff}, the TLR (black solid line) lies below both envelopes, suggesting that the baseline design satisfies the formal efficiency requirement (see Sect.~\ref{sec:req}).

\subsection{Signal to noise (S/N) ratio}\label{sec:SNR}
As for the throughput evaluation, we investigated two scenarios for the potential instrument performance in terms of the delivered S/N, by considering a set of reference parameters for the object spectrum, sky and observing conditions\footnote{Reference spectrum defined as $M_{\rm AB}$\,$=$\,18.77\,mag. The sky radiance and atmospheric transmission spectra were computed with the ESO {\sc skycalc} tool, assuming  new Moon, airmass\,$=$\,1.16 and a Precipitable Water Vapor of 30\,mm.}.

\begin{table}[h!]
\centering
\caption{Predicted S/N for different detector binning (spectral\,$\times$\,spatial) for the two performance scenarios for a 1\,hr exposure of a $U$\,$=$\,17.5\,mag. A0-type star at 313\,nm (0.007\,nm/bin).} 
\label{tab:m_d}
%====
\begin{tabular}{ccc}
\hline \hline
\textit{} & \textit{\begin{tabular}[c]{@{}c@{}}Goal\\ case\end{tabular}} & \textit{\begin{tabular}[c]{@{}c@{}}Conservative\\ case\end{tabular}} \\
\Xhline{1pt}
S/N (1\,$\times$\,1 binning) & 26 & 21 \\
S/N (1\,$\times$\,2 binning) & 28 & 23 \\ \hline
\end{tabular}
\end{table}

We first considered a `goal case', where the detector DC\,$=$\,0.5e$^-$/pix/hr (at 165\,K) and the efficiency of anti-reflection coatings, cement for doublets, slicer-mirror reflectivity and detector QE are set to the values from Table~\ref{tab:coat_perf} that have been assessed as feasible. Secondly, we investigated a more conservative case in which the detector DC\,$=$\,3e$^-$/pix/hr (at 173\,K) and all the efficiency values are taken from the manufacturers' official datasheets. The results are shown in Table~\ref{tab:m_d} which summarizes the trade-off performed for the HR mode.

These values were used to verify that the performance of the instrument baseline design would fulfill the requirements presented in Sect.~\ref{sec:req}.

\subsection{Image quality and resolving power}
In terms of image quality, spectral resolution and sampling, the two arms of the baseline design are expected to perform fully within the specifications from Sect.~\ref{sec:req}. 
Detailed ray-tracing of the design (Sect.~\ref{Optical_Design}) has demonstrated that the achievable optical point-spread-function (PSF) is below 1 pixel.

The effective resolving power has been computed from the FWHM Gaussian fit of the resulting line-spread-function (LSF), including all major blurring contributions: optical aberrations (optics PSF $=$ 0.65\,px for both arms), residual flexure (FWHM $=$ 0.5\,px) and signal diffusion in the detector (0.9\,px). The value of 0.5\,px for residual flexure is the maximum allocated movement of the spectra on the detector which will be achieved either by the stiffness of the mechanical structure or using the Active Flexure Compensation system.

The spectral resolving power ($R$) and spectral sampling ($Sx$) as a function of the imager slicer modes are given in Table~\ref{tab:m_res}.
These results confirm that the baseline design satisfies the requirements of the assembled science cases for $R$\,$\ge$\,20000, combined with good spectral sampling and a broad wavelength coverage from 300\,nm to the goal of $>$400\,nm.

\begin{table}[h!]
\centering
\caption{Resolving power ($R$) and spectral sampling ($Sx$) for the two slicer modes, i.e. different field of view and slitwidth. $Sx$ is computed as the FWHM Gaussian fit of the resulting line-spread-function (LSF) from the different optical blurring terms convolved with the geometrical sampling. The effective $R$ is then derived from the LSF FWHM.} 
\label{tab:m_res}
%====
\begin{tabular}{lcc}
\hline \hline
\textit{} & HR mode & LR mode \\ \Xhline{1pt}
FoV (arcsec) & 10\,$\times$\,1.5 & 10\,$\times$\,6 \\
Slitwidth (arcsec) & 0.25 & 1 \\
$R_{\rm mean}$ ($\lambda/\Delta\lambda$) & 24000 & 7200 \\
$R_{\rm min}$ ($\lambda/\Delta\lambda$) & 21000 & 6200 \\
$Sx_{\rm mean}$ (pix) & 2.43 & 8.35 \\
$Sx_{\rm min}$ (pix)  & 2.35 & 8.00 \\ \hline
\end{tabular}
\end{table}

\section{Model-Based Systems Engineering}
A Model-Based System Engineering (MBSE) approach \cite{Ref-MBSE} can help system engineers keep control of complex projects using the model as primary `Source of Truth' for all system engineering information related to the instrument. This approach requires the adoption of a tool and a modeling language (e.g. Cameo and SysML in our case \cite{Ref-CAMEO}). MBSE, however, faces difficult challenges in projects that are document-centric like those of ESO-related projects, therefore we also developed tools to produce the required documents from the model. 

Another challenge derives from the interaction with the various cultures of the engineering disciplines involved, namely: software, electronic, optical and mechanical engineers who are all used to different tools and languages to describe their domains. To cope with this and to avoiding imposing MBSE, which may look like an attempt to centralize and bureaucratize the project, we decided to keep the modeling environment restricted to the system-engineering team while interacting with the other engineers using traditional Excel files. This approach allowed us to both exploit the advantages of MBSE while not disturbing other engineering domains. Also, the Excel file can be used in syncing mode to allow data to be entered into the model; it should however be stressed that the model remains the only `Source of Truth'.

A fundamental role of MBSE is the requirements analysis. The TLRs from the ESO DOORS database \cite{Ref-DOORS} were imported in Cameo using the reqIf (Requirements Interchange Format) format. Analysis and flow-down of the requirements were then performed using SysML inside the model.

The use of MBSE in CUBES is laying the foundation for further applications of this approach in future astronomical projects. The experience accumulated in this project will be conveyed in a general SysML profile (AstroMBSE) that will be made available to the community.

\section{Conclusions}
We have presented the Phase A design of the Cassegrain U-Band Efficient Spectrograph (CUBES) for the VLT. Analysis of the design shows that it will deliver outstanding ($>$40\%) efficiency across its bandpass of 300--405\,nm, at a mean $R$\,$\sim$\,24000 (HR mode) and $R$\,$\sim$\,7000 (LR mode). We have also presented the option of a fiber link to UVES to provide the capability of simultaneous high-resolution spectroscopy at  $\lambda$\,$\ge$\,420\,nm.

With contributions from institutes in six countries, the CUBES design is well placed to deliver the most efficient ground-based spectrograph at near-UV wavelengths as we move into the construction phase, with science operations anticipated for 2028. Within this tight timeline, it is important to keep the system development under control by means of dedicated SysML profiles and procedures developed in the MBSE context. We highlighted the importance of these tools to maintain a high level of control and consistency in the instrument development.

\section*{Acknowledgments}
The INAF authors acknowledge financial support of the Italian Ministry of Education, University, and Research with PRIN 201278X4FL and the "Progetti Premiali" funding scheme.
The German authors acknowledge support from BMBF under grant 05A20VHA. R.S. acknowledges support by the National Science Centre, Poland, through project 2018/31/B/ST9/01469.

%\begin{acknowledgements}
%\end{acknowledgements}




\end{document}